\documentclass[nofootinbib,superscriptaddress,showkeys,showpacs,twocolumn,amsmath,amssymb]{revtex4-1}

\usepackage[T1]{fontenc}
\usepackage{aas_macros}

\newcommand{\abs}[1]{\left|#1\right|}
\newcommand{\vc}[1]{\boldsymbol{#1}}
\newcommand{\ord}[1]{{}^{(#1)}}
\newcommand{\ocn}[1]{{\mathcal{O}\left(c^{-#1}\right)}}
\newcommand{\dd}{\mathrm{d}}

\newcommand{\Sec}[1]{Sec.~\ref{#1}}
\newcommand{\eqn}[1]{Eq.~\eqref{#1}}
\newcommand{\eqns}[1]{Eqs.~\eqref{#1}}

\begin{document}

\title{Probing the dark matter issue in $f(R)$-gravity via gravitational lensing}

\author{M.~Lubini}
\email{lubini@physik.uzh.ch}
\affiliation{Institut f\"{u}r Theoretische Physik, Universit\"{a}t Z\"{u}rich, Winterthurerstrasse 190, CH-8057 Z\"{u}rich, Switzerland}

\author{C.~Tortora}
\affiliation{Institut f\"{u}r Theoretische Physik, Universit\"{a}t Z\"{u}rich, Winterthurerstrasse 190, CH-8057 Z\"{u}rich, Switzerland}

\author{J.~N\"af}
\affiliation{Institut f\"{u}r Theoretische Physik, Universit\"{a}t Z\"{u}rich, Winterthurerstrasse 190, CH-8057 Z\"{u}rich, Switzerland}

\author{Ph.~Jetzer}
\affiliation{Institut f\"{u}r Theoretische Physik, Universit\"{a}t Z\"{u}rich, Winterthurerstrasse 190, CH-8057 Z\"{u}rich, Switzerland}

\author{S.~Capozziello}
\affiliation{Dipartimento di Scienze Fisiche, Universit\`{a} degli studi di Napoli ``Federico II'', and INFN, Sezione di Napoli Complesso Universitario di Monte S.~Angelo, Via Cinthia, Edificio N, 80126 Napoli, Italy}



\begin{abstract}
For a general class of analytic $f(R)$-gravity theories, we discuss the weak field limit in view of gravitational lensing.
Though an additional Yukawa term in the gravitational potential modifies dynamics with respect to the standard Newtonian limit of General Relativity, the motion of massless particles results unaffected thanks to suitable cancellations in the post-Newtonian limit.
Thus, all the lensing observables are equal to the ones known from General Relativity.
Since $f(R)$-gravity is claimed, among other things, to be a possible solution to overcome for the need of dark matter in virialized systems, we discuss the impact of our results on the dynamical and gravitational lensing analyses.
In this framework, dynamics could, in principle, be able to reproduce the astrophysical observations without recurring to dark matter, but in the case of gravitational lensing we find that dark matter is an unavoidable ingredient.
Another important implication is that gravitational lensing, in the post-Newtonian limit, is not able to constrain these extended theories, since their predictions do not differ from General Relativity.
\end{abstract}

\maketitle


\section{Introduction}\label{sec:introduction}

Einstein's General Relativity (GR) has been extensively adopted to describe a wide set of astrophysical observations.
On the one hand, at scales of the Solar System, it has been shown to reproduce very accurately many precise astrophysical measurements, such as the gravitational bending of light, the perihelion precession of Mercury, the Shapiro time delay and the mass diagrams of double pulsars \cite{Burgay06, Lyne06, Will06}.
On the other hand, at larger scales, it has required the introduction of some unknown and unseen sources, i.e.~dark matter and dark energy, to reproduce, e.g.~the missing mass in galaxies and galaxy clusters and the accelerate rate of expansion of the Universe \cite{Carroll92, Sofue01, Komatsu11}.
The existence of these energy sources has not yet been proved neither through direct detections nor with indirect observations.
However, in some clusters of galaxies significant displacements between the lensing and the baryonic mass give an indirect hint for the existence of dark matter \cite{Clowe06, Bradac08, Sereno10}.

In order to overcome these shortcomings affecting Einstein's predictions, alternative theories of gravity have been proposed.
An empirical approach to get over the need of dark matter in galaxies is, for example, given by the MOdified Newtonian Dynamics (MOND), which relies on a modification of Newton's second law reproducing the observed galaxy dynamics \cite{Sanders02, Cardone11mond}.
In the recent years, this theory has received a renewed interest after the proposal of a possible fully covariant relativistic formulation referred to as Tensor-Vector-Scalar gravity (TeVeS) \cite{Bekenstein04}.

More elegant and theoretically motivated ways can be obtained by straightforward generalizations of GR.
Among such extended theories, particular attention has been devoted to the so called $f(R)$-gravity.
This is a fourth-order formulation of the field equations in the metric approach, based on a generalization of the Einstein-Hilbert Lagrangian to nonlinear functions $f(R)$ of the Ricci scalar.
For a comprehensive overview we refer to \cite{Nojiri07, Capozziello08a, Capozziello09e, deFelice10, Capozziello10, Capozziello11, Nojiri11}.

In order to be consistent, $f(R)$-gravity needs to reproduce observations going from Solar System up to cosmological scales.
Since it turned out that they induce modifications in the gravitational potential already in the weak field limit, they modify the dynamics at local scales too \cite{Stelle78, Capozziello04, Naef10}.
Thus, as these theories have to reproduce also the Solar System tests, constraints have been obtained from different measurements, as the E\"{o}t-Wash experiment \cite{Kapner07}, the geodesic precession of gyroscopes measured by Gravity Probe B and the precession of the binary pulsars \mbox{PSR J0737-3039} \cite{Naef10}.

These extended theories of gravity have first been proposed as candidates to replace dark energy \cite{Capozziello02, Nojiri03b, Capozziello03, Nojiri03}, and more recently they have also been adopted to reproduce the dynamical observations in galaxies and cluster of galaxies without the presence of large amounts of dark matter
needed in GR \cite{Sanders02, Capozziello04, Capozziello06, Capozziello07, Frigerio07, Mendoza07, Sobouti07, Capozziello09b}.
In order to get rid of the presence of dark matter, $f(R)$-gravity has to consistently reproduce not only the dynamical observations, but also other mass probes coming from e.g.~gravitational lensing.

In this paper, we investigate the equation for the light deflection considering the weak field limit of a general class of $f(R)$-gravity.
We particularly pay attention to the possible implications which gravitational lensing has on the attempt to avoid the presence of dark matter.

In \Sec{sec:theory} we introduce $f(R)$-gravity and its field equations, whereas \Sec{sec:potentials} is devoted to the expansion and the resulting solutions for the metric elements.
The implications on particle dynamics are discussed \Sec{sec:particle}, while in \Sec{sec:lensing} we derive the deflection angle and discuss the gravitational lensing phenomena thereof arising.
Finally, in \Sec{sec:discussion} we discuss the results obtained, with particular care to the role of the gravitational potentials in lensing and dynamical observables, and we give some future prospects.

Throughout the paper the metric tensor $g_{\mu\nu}$ has signature $(-,+,+,+)$.


\section{$f(R)$-gravity}\label{sec:theory}

$f(R)$-gravity considers a generalized gravity Lagrangian, which in the GR case reduces to the Ricci scalar.
These theories have a modified Einstein-Hilbert action that reads
\begin{equation}\label{eqn:action}
	S = \frac{1}{2\mathcal{X}}\int f(R)\sqrt{-g}\,\dd^4x + S_M,
\end{equation}
where $g$ is the determinant of the metric tensor $g_{\mu\nu}$, \mbox{$\mathcal{X}=8\pi G c^{-4}$}, $R = g^{\mu\nu}R_{\mu\nu}$ is the Ricci scalar and $f(R)$ is some general function of $R$.

The variation of action \eqref{eqn:action} with respect to the metric $g_{\mu\nu}$ yields the Euler-Lagrange equations
\begin{equation}\label{eqn:e-l}
	f'(R)R_{\mu\nu} - \frac{f(R)}{2}g_{\mu\nu} - \nabla_{\mu}\nabla_{\nu}f'(R) + g_{\mu\nu}\square f'(R) = \mathcal{X}\,T_{\mu\nu},
\end{equation}
where $\nabla_{\mu}$ is the covariant derivative for the metric, \mbox{$\square = \nabla^{\mu}\nabla_{\mu}$} and $T_{\mu\nu} = (-2/\sqrt{-g})(\delta S_M/\delta g^{\mu\nu})$ is the energy-momentum tensor.

By taking the trace of \eqref{eqn:e-l}, one obtains
\begin{equation}\label{eqn:trace}
	3\square_g f'(R) + f'(R)R - 2f(R) = \mathcal{X}\,T,
\end{equation}
where $T$ is the trace of $T_{\mu\nu}$.
Eq. \eqref{eqn:trace} describes the evolution of the Ricci scalar as a dynamical quantity.
Using this equation, we can rewrite \eqn{eqn:e-l} as
\begin{equation}\label{eqn:Rmn}
\begin{split}
	R_{\mu\nu} & = \frac{1}{f'(R)}\left[\mathcal{X}\left(T_{\mu\nu}-\frac{1}{3}g_{\mu\nu}T\right) -\frac{1}{6}g_{\mu\nu}f(R)\right.\\
	&\quad \left. {}+ \frac{1}{3}g_{\mu\nu}f'(R)R + \nabla_{\mu}\nabla_{\nu}f'(R)\right],
\end{split}
\end{equation}
where the function $f'(R)$ has to be considered as an effective scalar field representing the further gravitational degrees of freedom of the theory \cite{Capozziello08b, Bellucci09, Bogdanos10}. In fact, it is worth stressing that trace equation \eqref{eqn:trace} is a Klein-Gordon equation, where $\phi=f'(R)$ and $V'(\phi)=[f'(R)R - 2f(R)]/3$ \cite{Nojiri07, Capozziello08a, Capozziello09e, deFelice10, Capozziello10, Capozziello11}.


\section{Gravitational potentials}\label{sec:potentials}

\subsection{The $1/c$ expansion}

The basic assumption in deriving the equations for gravitational lensing is that the gravitational field is weak and stationary.
Furthermore, for most astrophysical phenomena included gravitational lensing, the energy-momentum tensor is well approximated by a perfect fluid with mass density $\rho$ and pressure $p$ \cite{Weinberg72}. Thus the assumptions are that the Newtonian potential $\Phi$ of the mass distribution, the typical velocities $v$ and the pressure of the fluid obey $\abs{\Phi} \ll c^{2}$, $v\ll c$ and $\abs{p}\ll\rho\,c^2$, respectively.
Accordingly, we can treat the equations in a perturbative way, by expanding them in inverse powers of the speed of light as done in \cite{Capozziello07b, Clifton08, Naef10, Stabile10}.
This expansion as for the parameterized post-Newtonian (PPN) formalism is defined by
\begin{equation}\label{eqn:ord-of-small}
	\frac{\Phi}{c^2}\sim\frac{v^2}{c^2}\sim\frac{p}{\rho\, c^2}\sim\Pi\sim\ocn{2},
\end{equation}
where $\Pi$ is the ratio of the energy density to the rest-mass density.

In addition, we also assume that in absence of a gravitational field the background space-time is flat.
This assumption implies that $f(0)=0$, which is equivalent to neglect the contribution of a possible cosmological constant to the $f(R)$-gravity.
In a weak field regime the metric tensor can thus be expanded about the Minkowski tensor $\eta_{\mu\nu}$
\begin{equation}\label{eqn:g-lin}
	g_{\mu\nu}=\eta_{\mu\nu}+h_{\mu\nu},
\end{equation}
where $\abs{h_{\mu\nu}} \ll 1$.

One can see from the equation of motion to which order the components of the metric perturbation $h_{\mu\nu}$ have to be taken into account.
For timelike particles propagating along a geodesic, at the leading order of the expansion, only the knowledge of $\ord{2}h_{00}$\footnote{We denote with $\ord{n}$ terms of order $\ocn{n}$.} is required.
This corresponds to the Newtonian limit.
In this limit, for null particles only the background metric $\eta_{\mu\nu}$ is needed, which implies that the trajectories are straight lines.
Accordingly, for photons there is no light deflection and thus we need the post-Newtonian limit, where at the leading order the knowledge of both $\ord{2}h_{00}$ and $\ord{2}h_{ij}$ is required \cite{Clifton08}.
For our purpose it is thus enough to consider terms up to $\ocn{2}$.
At this order the components of the metric tensor can be written as
\begin{align}
\begin{split}\label{eqn:linearization}
	g_{00} & = -1 + \ord{2}h_{00} + \ocn{4} \\
	g_{0i} & = \ocn{3} \\
	g_{ij} & = \delta_{ij} + \ord{2}h_{ij} + \ocn{4} .
\end{split}
\intertext{Using the expansion \eqref{eqn:linearization} we can than compute the components of the Ricci tensor}
\begin{split}\label{eqn:Rmn(2)}
	R_{00} & = -\frac{1}{2}\nabla^2\ord{2}h_{00} + \ocn{4} \\
	R_{0i} & = \ocn{3} \\
	R_{ij} & = \frac{1}{2}\left(-\nabla^2\;\ord{2}h_{ij} + \ord{2}h_{00,ij} - \ord{2}h_{kk,ij}\right. \\
	&\quad\left.{} + \ord{2}h_{ik,kj} + \ord{2}h_{kj,ki} \right) + \ocn{4} ,
\end{split}
\intertext{while, assuming $f(R)$ to be analytic at $R=0$, to the same order the Ricci scalar and consequently $f(R)$ and $f'(R)$ read}
\begin{split}
	R & = \ord{2}R + \ocn{4}\\
	f(R) & = f'(0) \ord{2}R + \ocn{4} \\
	f'(R) & = f'(0) + f''(0) \ord{2}R + \ocn{4} .
\end{split}
\intertext{The same expansion has to be done for the components of the energy-momentum tensor, which to the leading order are}
\begin{split}
	\mathcal{X}\,T^{00} & = \mathcal{X}\ord{-2}T^{00} + \ocn{4} \\
	\mathcal{X}\,T^{0i} & = \ocn{3} \\
	\mathcal{X}\,T^{ij} & = \ocn{4}.
\end{split}
\end{align}
Hereafter $f'(0)$ is assumed to be different from $0$, in order to deal with these $f(R)$ theories as a modification of Einstein's theory and thus do not exclude GR as a particular case. In this approach, GR is nothing else but the first term of the Taylor expansion of a more general theory. This means that $f(R)$-gravity is a class of extended theories of gravity \cite{Nojiri07, Capozziello08a, Capozziello09e, deFelice10, Capozziello10, Capozziello11}.
We can also assume without loss of generality that $f''(0)\neq0$, because otherwise at order $\ocn{2}$ \eqn{eqn:trace} yields $\ord{2}R=-\mathcal{X}\,\ord{2}T_{00}/f'(0)$ and $f'(R)$ is constant, which is nothing else than the GR case.

At this point it is important to discuss and to understand which are the hidden assumptions the previous expansions are bringing with.
Since the expansion \eqref{eqn:ord-of-small} deals with dimensionless quantities, we have to be careful when expanding quantities that carry physical dimensions.
The constant $f''(0)$ carries dimension length$^{2}$, thus the whole term $f''(0)\ord{2}R$, which is dimensionless has to be of order $\ocn{2}$ and respectively all other terms have to be of higher order.
From a mathematical point of view, where expansions are done among infinitesimally small quantities, this might be true, as the expansion coefficients either vanish or their absolute values are strictly grater than zero.
Whereas in physical cases, when we are dealing with small but not infinitesimally small quantities, this might not be true and can thus lead to some issues.
This expansion thus breaks down if $\ord{2}R\gtrsim f'(0)/f''(0)$. In other words, this means that we have to confront our approximation with an effective mass where the post-Newtonian limit could not work (for details see \cite{Capozziello08b, Bellucci09, Bogdanos10}).

\subsection{The Metric elements}

In order to simplify the metric tensor, without loss of generality due to gauge freedom, we choose the coordinates such that the three gauge conditions \cite{Clifton08, Naef10}
\begin{equation}\label{eqn:gauge}
	g_{ij,j} - \frac{1}{2}\left(g_{jj}-g_{00}\right)_{,i} - \frac{f'(R)_{,i}}{f'(R)} = \ocn{4}
\end{equation}
are satisfied. As a consequence of this choice the spatial part of the metric tensor is diagonal.

Using condition \eqref{eqn:gauge} and the expansions obtained through \eqn{eqn:linearization}, the \eqns{eqn:trace} and \eqref{eqn:Rmn} at order $\ocn{2}$ reduce to the following set of equations
\begin{align}
	\nabla^2\ord{2}R & = \frac{1}{3f''(0)}\mathcal{X}\ord{-2}T^{00} + \frac{f'(0)}{3f''(0)}\ord{2}R \label{eqn:2R} \\
	\nabla^2\ord{2}h_{00} & = - \frac{\mathcal{X}}{f'(0)}\ord{-2}T^{00} + \frac{f''(0)}{f'(0)}\nabla^2\ord{2}R \label{eqn:2h00} \\
	\nabla^2\ord{2}h_{ij} & = - \left(\frac{\mathcal{X}}{f'(0)}\ord{-2}T^{00} + \frac{f''(0)}{f'(0)}\nabla^2\ord{2}R\right)\delta_{ij} \label{eqn:2hij}.
\end{align}
\eqns{eqn:2h00} and \eqref{eqn:2hij} have been obtained from the components of \eqn{eqn:Rmn}, where $\ord{2}R$ has been substituted by means of \eqn{eqn:2R}.

For simplicity hereafter we renormalize the gravitational constant $G_{\text{new}}=G_{\text{old}}/f'(0)$ and define $a^2:=f'(0)/(3f''(0))$.
The parameter $a$ can be assumed real, because only in this case the model is physically viable in agreement with the Cauchy problem which, in this case, is well formulated and well posed \cite{Capozziello09c}.
In addition, the gravitational source is assumed to be a perfect fluid, where the $0\!-\!0$ component of the energy-momentum tensor at order $\ocn{2}$ is given by $\ord{-2}T^{00}=\rho\,c^2$.

\eqn{eqn:2R} is an inhomogeneous Helmholtz equation, which has the solution \cite{Clifton08, Naef10}
\begin{equation}\label{eqn:s2R}
	\ord{2}R = -\frac{2a^2}{c^2}\Psi(\vc{x},t) ,
\end{equation}
with the Yukawa potential
\begin{equation}\label{eqn:Psi}
	\Psi(\vc{x},t) := - G \int\frac{\rho(\vc{x}',t) e^{-a\left|\vc{x}-\vc{x}'\right|}}{\left|\vc{x}-\vc{x}'\right|}\dd^3x',
\end{equation}
whereas \eqns{eqn:2h00} and \eqref{eqn:2hij} are inhomogeneous Poisson equations and their solutions are
\begin{align}
	\ord{2}h_{00}(\vc{x},t) & = - \frac{2}{c^2}\left(\Phi(\vc{x},t) + \frac13 \Psi(\vc{x},t)\right) \label{eqn:s2h00}\\
	\ord{2}h_{ij}(\vc{x},t) & = - \frac{2}{c^2} \left(\Phi(\vc{x},t) - \frac13 \Psi(\vc{x},t)\right) \delta_{ij}, \label{eqn:s2hij}
\end{align}
with the Newtonian potential
\begin{equation}\label{eqn:Phi}
	\Phi(\vc{x},t) := - G \int\frac{\rho(\vc{x}',t)}{\left|\vc{x}-\vc{x}'\right|}\dd^3x'.
\end{equation}

The GR limit for $f(R)$-gravity in a weak field is given by $a\to\infty$, which corresponds to $f''(0)\to0$.
In this limit the Newtonian potential $\Phi$ remains unchanged, while the Yukawa potential $\Psi$ vanishes exponentially.
For the metric components \eqref{eqn:s2h00} and \eqref{eqn:s2hij} we then obtain the GR limit, i.e.~$\ord{2}h_{00}=\ord{2}h_{ii}=-2\Phi/c^2$, as expected.

These equations are valid for continuous mass distributions.
In the particular case of a point source with mass $m$ fixed at the origin of the coordinates $\vc{x}=0$, the mass density is $\rho(\vc{x},t)=m\,\delta^3(\vc{x})$ and the potentials reduce to the simpler formulae
\begin{align}\label{pot1}
	\Phi(\vc{x}) & = -\frac{G\,m}{\abs{\vc{x}}} \\
	\Psi(\vc{x}) & = -\frac{G\,m\,e^{-a\abs{\vc{x}}}}{\abs{\vc{x}}}.
\end{align}


\section{Particle dynamics}\label{sec:particle}

After having determined explicit expressions for the metric elements in terms of the gravitational potential $\Phi$ and $\Psi$, we shortly discuss the equations of motion for massive particles within the weak field limit of $f(R)$-gravity.
The geodesic equation
\begin{equation}
	\frac{\dd^2x^\alpha}{\dd\tau^2}=-\Gamma^\alpha_{\beta\gamma}\frac{\dd x^\beta}{\dd\tau}\frac{\dd x^\gamma}{\dd\tau}
\end{equation}
in the Newtonian limit for massive particles, which is the leading order, reads
\begin{equation}\label{eqn:vdot}
	\frac{\dd\vc{v}}{\dd t}=-\frac{c^2}{2}\vc{\nabla}h_{00}=\vc{\nabla}\left(\Phi+\frac13\Psi\right),
\end{equation}
where $\vc{v}=\frac{\dd\vc{x}}{\dd t}$ and, as we have already pointed out previously, only the $0\!-\!0$ component of the metric tensor is needed.
Hence, the equation of motion contains the additional Yukawa potential $\Psi$, which modifies the dynamics already at the Newtonian limit.

In the particular case of a spherically symmetric mass distribution centered at the origin $\vc{x}=0$, the potentials only depend on the radial distance to the center $r=\abs{\vc{x}}$.
Then, particles moving on circles have the circular velocity
\begin{equation}\label{eqn:vc}
	v_c(r)=\sqrt{\frac{GM(r)}{r}+\frac{r}{3}\frac{\partial\Psi(r)}{\partial r}}.
\end{equation}
$v_c$ is an important observed quantity in systems with an ordered motion.
In the case of a thin disk, which is the standard approximation for spiral galaxies, the circular velocity is given by $v_c(R)^2=R(\frac{\partial}{\partial R}\Phi+\frac13\frac{\partial}{\partial R}\Psi)$, with $R$ the distance to the center of the disk.

The dynamics in kinematically hot systems, like early-type galaxies and clusters of galaxies is instead not ordered but random.
A relevant quantity for those systems is the velocity dispersion of objects like stars in a galaxy or galaxies in a cluster.
Starting from the collisionless Boltzmann equation, assuming spherical symmetry and no rotation, the radial Jeans equation can be derived \cite{Binney87}.
Under the hypothesis of isotropy, the radial velocity dispersion is given by
\begin{equation}\label{eq:iso}
	\sigma_r^2 (r) = \frac{1}{\rho(r)} \int_r^\infty \rho(r') \left( \frac{GM(r')}{r'} + \frac13\frac{\partial \Psi(r')}{\partial r'} \right) \dd r' ,
\end{equation}
where $\rho(r)$ is the spherically symmetric mass density profile\footnote{The radial velocity dispersion needs to be projected along the line of sight and within circular or rectangular apertures in order to be compared with the spectral observations \cite{Tortora09}.}.


\section{Gravitational Lensing}\label{sec:lensing}

In principle, when compared with GR, f(R)-gravity could have not only a different impact on the dynamics of massive particles, but also on the bending of light rays \cite{Capozziello06b}.
In this section, we work out the deflection angle for light rays in a gravitational field, whose derivation, together with all the gravitational lensing phenomena arising out from it, is equivalent to GR \cite{Schneider92}.

As stated in \Sec{sec:theory} the basic assumption in gravitational lensing is that the gravitational field is weak and stationary.
Using \eqns{eqn:s2h00} and \eqref{eqn:s2hij}, the metric can thus be expanded and written as
\cite{Weinberg72, Schneider92},
\begin{equation}\label{eqn:ds2}
	\mathrm{d}s^2=-(1-\ord{2}h_{00})c^2\mathrm{d}t^2+(\delta_{ij}+\ord{2}h_{ij})\mathrm{d}x^i\mathrm{d}x^j+\ocn{3}.
\end{equation}
Since light rays propagate along null geodesics, $\dd s^2=0$ and
consequently keeping only terms up to second order $\ocn{2}$, \eqn{eqn:ds2} reduces to
\begin{equation}\label{eqn:cdt2}
	c^2\mathrm{d}t^2=\frac{\delta_{ij}+\ord{2}h_{ij}}{1-\ord{2}h_{00}}\mathrm{d}x^i\mathrm{d}x^j=\frac{1+\ord{2}h}{1-\ord{2}h_{00}}\dd l^2,
\end{equation}
where $h$ is defined such that \eqn{eqn:s2hij} can be written as $\ord{2}h_{ij}=\ord{2}h\,\delta_{ij}$ and $\dd l=\abs{\dd\vc{x}}$ is the Euclidean arc length.

Fermat's principle states that a light ray from a source $\mathcal{S}$ to an observer $\mathcal{O}$, that moves along a world line, follows a null curve $\gamma$ that extremizes the
observer's arrival time $\tau$ under variation of $\gamma$, meaning that $\delta\tau=0$.
Considering a stationary observer, i.e.~choosing the coordinates such that the observer's world line has constant $x^i$s, the arrival time $\tau=t$, except for a constant, is given by
\begin{equation}\label{eqn:arr_time}
	t = \frac1c \int_{\tilde\gamma} n \,\dd l,
\end{equation}
where $\tilde\gamma$ is the spatial projection of the null geodesic and $n$ is considered as an effective index of refraction, defined by
means of \eqn{eqn:cdt2} as
\begin{equation}
	n=\sqrt{\frac{1+\ord{2}h}{1-\ord{2}h_{00}}},
\end{equation}
which at order $\ocn{2}$ reduces to
\begin{equation}
	n = 1+\frac12\left(\ord{2}h_{00}+\ord{2}h\right).
\end{equation}
Therefore for the metric components \eqref{eqn:s2h00} and \eqref{eqn:s2hij} we obtain
\begin{equation}
	n = 1-\frac{2}{c^{2}}\Phi.
\end{equation}
This effective refraction index contains only the Newtonian potential $\Phi$ and not the Yukawa potential $\Psi$.
In fact, it is exactly the same as in GR.
The reason is that because of the symmetry between the metric components \eqref{eqn:s2h00} and \eqref{eqn:s2hij}, the potential $\Psi$ cancels out at order $\ocn{2}$ and appears only at higher order terms.
This implies that the deflection angle and all the gravitational lensing phenomena arising out from it are not distinguishable from the GR ones.
While for massive particles $f(R)$-gravity, in the weak field regime, can produce different equations of motion, in the case of photons these theories are equivalent to GR. Possible deviations could be revealed only in stronger regimes, that is at approximation orders higher than $\ocn{2}$ but not at post-Newtonian limit.

From the variation of the arrival time in \eqn{eqn:arr_time}, i.e.~$\delta t=0$, one can obtain
the deflection angle \cite{Schneider92}
\begin{equation}\label{eqn:angle}
	\hat{\vc{\alpha}} = \hat{\vc{\alpha}}_{\mathrm{GR}} = \frac{2}{c^{2}}\int \vc{\nabla}_{\perp} \Phi \,\dd l,
\end{equation}
where $\vc{\nabla}_{\perp} := \vc{\nabla} - \hat{\vc{e}}(\hat{\vc{e}}\cdot\vc{\nabla})$, with $\hat{\vc{e}}:=\frac{\dd\vc{x}}{\dd l}$ the unit spatial vector tangent to the direction of the light ray.
For an axially symmetric mass distribution around the line of sight the deflection angle is given by the formula
\begin{equation}
	\hat{\alpha}(\xi)=\frac{4GM_{\text{cyl}}(\xi)}{c^2\xi},
\end{equation}
where $\xi$ is the distance from the line of sight and $M_{\text{cyl}}(\xi)$ is the mass enclosed in a cylinder with radius $\xi$.

Some recent works have suggested that when the spherically symmetric mass distribution $\rho(r)$ is left very general, in the weak field limit
of $f(R)$-gravity, the gravitational potential assumes a very general form in spherical coordinates \cite{Faulkner07, Capozziello09d, Cardone11yukawa}
\begin{equation}\label{eqn:metric_delta}
\begin{split}
	\dd s^{2} & = - \left ( 1- \frac{G M}{r} - \frac{\delta \, e^{ - a r}}{3 a r} \right )c^{2} \dd t^{2} \\
	&\quad + \left ( 1 + \frac{G M}{r} - \frac{\delta \, (r a + 1) ~ e^{ - a r}}{3 a r} \right ) \dd r^{2} + r^{2} \dd\Omega,
\end{split}
\end{equation}
where $\delta$ and $a$ are taken as two free parameters of the theory\footnote{We notice that for a point mass the metric elements in \eqn{eqn:metric_delta} are equivalent to \eqns{eqn:s2h00} and \eqref{eqn:s2hij} with $\delta=a$, if one passes from spherical to Cartesian coordinates.}.
The Chameleon effect takes place, leading to the above metric, where the new parameter $\delta$ is shown to depend on both the source mass and the local density which it is embedded in \cite{Faulkner07, Cardone11yukawa}. Clearly, in this regime, the validity of the Birkhoff and Gauss theorem is not guaranteed and spherically symmetric solutions could be achieved also for the Ricci scalar depending on the radius \cite{Capozziello07b}. It is worthwhile noticing that, in any case, conservation laws hold thanks to the Bianchi identities.

However, our conclusions about gravitational lensing predictions remain valid also for the metric \eqref{eqn:metric_delta}.
The reason once again is that the additional potentials, which in this case depend on the parameters $a$ and $\delta$, exactly cancel out in the effective refraction index.


\section{Discussion and Conclusion} \label{sec:discussion}

In the present paper we discuss the weak field limit for a very general class of $f(R)$-gravity theories.
In most astrophysical phenomena from the typical Solar System observations, to the dynamics in galaxies and clusters, as well as gravitational lensing observations, the gravitational field generated by the respective mass distribution can be considered as weak, allowing the field equations to be treated in a perturbative way \cite{Clifton08, Naef10, Stabile10}.
From the consequent expansion in inverse powers of the speed of light, the time-time and space-space metric elements deliver the gravitational potentials of the mass distribution that govern the motion of both massive and massless particles.
While on the one side, within the Einsteinian framework, only the classical Newtonian $\Phi(r)$ potential is recovered, on the other side, in the case of the more general $f(R)$-gravity theories, the additional Yukawa potential $\Psi(r)$ emerges.
The equations of motion of massive particles, and in particular, the dynamical observables like the circular velocity and the velocity dispersion, differ from the Einsteinian ones, as they depend on a combination of both potentials $\Phi$ and $\Psi$.
The remarkable result here obtained is that, despite massive particles are affected by such an extension of the gravity theory, massless particles are not, since they are governed only by the Newtonian potential.
{\it Therefore, all the gravitational lensing predictions obtained from GR are still valid in the framework of $f(R)$-gravity.}

Our results have been obtained taking into account only terms up to the leading order $\ocn{2}$.
For most astrophysical phenomena the higher order terms can be neglected.
In fact, the weak field conditions \eqref{eqn:ord-of-small} hold for stars, galaxies and clusters, which induce slight perturbations of the space-time.
Consequently the derived \eqn{eqn:vdot} for particle dynamics and \eqn{eqn:angle} for gravitational lensing are not more valid in the case of sources that strongly act on space-time, like neutron stars and black holes.
To appropriately describe these cases, the metric elements need to be expanded to higher orders, hence gravitational lensing phenomena in $f(R)$-gravity can differ from GR in the strong field regime.

The dark components account for $\sim 95 \%$ of the total energy budget of the Universe \cite{Komatsu11} and their existence has not yet been satisfyingly proved.
Extended theories of gravity have been proposed as candidates to replace not only the dark energy on cosmological scales \cite{Capozziello02, Capozziello03, Capozziello08a}, but also to account for the observed mass profiles in galaxies and clusters of galaxies without the inclusion of dark matter \cite{Sanders02, Capozziello04, Capozziello06, Capozziello07, Frigerio07, Mendoza07, Sobouti07, Capozziello09b}.

The need of dark matter in astrophysical objects emerges from different observational mass probes.
In galaxies both lensing observations \cite{Auger09, Tortora10} and dynamical analyses, such as circular rotation curves in spiral galaxies \cite{Sofue01} and velocity dispersions in early-type systems \cite{Napolitano09, Romanowsky09, Tortora09} have showed that dark matter is the dominant mass component not only in the outskirts but also in the very central regions.
The same is found in galaxy clusters, where a large amount of dark matter is needed to explain the weak and strong lensing observations \cite{Clowe06, Bradac08, Sereno10} as well as dynamical data like the velocity dispersion of the galaxies or the temperature profile of the X-ray emitting intracluster medium \cite{deFilippis04, deFilippis05}.
In these analyses almost the same amount of dark matter is required in both dynamical and lensing data.

Up to now, $f(R)$-gravity has been mainly tested using dynamical data in both spiral galaxies \cite{Capozziello07, Frigerio07} and galaxy clusters \cite{Capozziello09a}.
Since the dynamical predictions of the here discussed $f(R)$ theories depend on the additional Yukawa potential (as shown in \eqns{eqn:vc} and \eqref{eq:iso}), the observed data could, in principle, be reproduced without the need of dark matter by fitting the free parameter $a$ of the theory.

However, gravitational lensing represents a complementary mass probe to determine if $f(R)$-gravity theories are valuable ways to escape the presence of dark matter.
In the case of the here discussed $f(R)$ theories, we have shown that the gravitational lensing observables are the same as the GR ones.
Since in the GR framework the baryonic matter in galaxies and cluster of galaxies is not enough to explain the lensing observations, in the regime of these $f(R)$-gravity theories this is the case too.
{\it Therefore, from the viewpoint of gravitational lensing, dark matter is still needed in $f(R)$-gravity to fit the observations.}

However, an important remark is due at this point.
Clearly, dark matter shows particle properties that could be related to the further degrees of freedom of $f(R)$-gravity.
In this case, there is no contradiction with the above results since null-geodesics, in the post-Newtonian approximation, follow the same behavior of GR while particle dynamics is different (see \cite{Nojiri08} for a detailed discussion).

Another important implication is that gravitational lensing is not able to constrain the parameter space of the here assumed $f(R)$-gravity.
While from precise Solar System tests, constraints on the parameter $a$ have been given via dynamical predictions \cite{Kapner07, Naef10}, gravitational lensing predictions from $f(R)$-gravity are in agreement with the observations, as they do not differ from the GR ones.
Following a similar approach, recently \citet{Berry11} have found that these $f(R)$-gravity theories have an effective post-Newtonian parameter $\gamma=1$, confirming our results.

Combining the dynamical and lensing mass probes, as they have to be coherent, one would expect to be able to fit at most a small part of the missing mass, and still need a non negligible amount of dark matter, which a modification of the theory, like the one here discussed, is not able to avoid.
In a forthcoming paper, we plan to investigate further on this supposition, performing a combined dynamical and lensing analysis in lens galaxies \cite{Auger09, Tortora10}.
Moreover it would be interesting to analyse more general $f(R)$-gravity theories, for example theories with a nonvanishing cosmological constant, i.e.~$f(0)\neq0$, and their implication on dynamical and lensing observables.

A final remark concerns the ``representation'' of $f(R)$-gravity.
According to \cite{Capozziello06c}, this kind of theories can be recast as perfect fluids or as scalar-tensor theories.
This means that dark matter (and dark energy) assumes different dynamics depending on the frame which can be the Jordan or the Einstein frame.
Gravitational lensing approximation has to be dealt accordingly taking care of how the gravitational degrees of freedom are represented in the two different frames.
In the discussion performed in this paper, we considered only the Jordan frame without assuming any conformal transformation.


\begin{acknowledgments}
CT is supported by the Swiss National Science Foundation.
\end{acknowledgments}


\bibliography{bibliography}

\end{document}